# Hybrid Surface Patterns Mimicking the Design of the Adhesive Toe Pad of Tree Frog

Longjian Xue[*,†,‡,§], Belén Sanz[§,▲], Aoyi Luo[#], Kevin T. Turner[#], Xin Wang[‡], Di Tan[‡], Rui Zhang[‡], Hang Du[‡], Martin Steinhart[¶], Carmen Mijangos[▲], Markus Guttmann[Δ], Michael Kappl[§], and Aránzazu del Campo[*,§,⊥,∥]

[†]The Institute of Technological Science and [‡]School of Power and Mechanical Engineering, Wuhan University, South Donghu Road 8, Wuhan 430072, China

[§] Max-Planck-Institut für Polymerforschung, Ackermannweg 10, 55128 Mainz, Germany

[⊥] INM − Leibniz Institute for New Materials, Campus D2 2, 66123 Saarbrücken, Germany
[∥] Chemistry Department, Saarland University, 66123 Saarbrücken, Germany

[¶] Institut für Chemie neuer Materialien, Universität Osnabrück, Barbarastr. 7, 49069 Osnabrück, Germany

[#] Department of Mechanical Engineering and Applied Mechanics, University of Pennsylvania, 220 S. 33rd Street, Philadelphia, Pennsylvania 19104-6315, United States

[Δ] Institute of Microstructure Technology, Karlsruhe Institute of Technology, Hermann-von-Helmholtz-Platz 1, 76344 Eggenstein-Leopoldshafen, Germany

[▲] Instituto de Ciencia y Tecnología de Polímeros, Consejo Superior de Investigaciones Científicas (CSIC), Juan de la Cierva 3, 28006 Madrid, Spain

*E-mail for L.X.: xuelongjian@whu.edu.cn
*E-mail for A.d.C.: aranzazu.delcampo@leibniz-inm.de

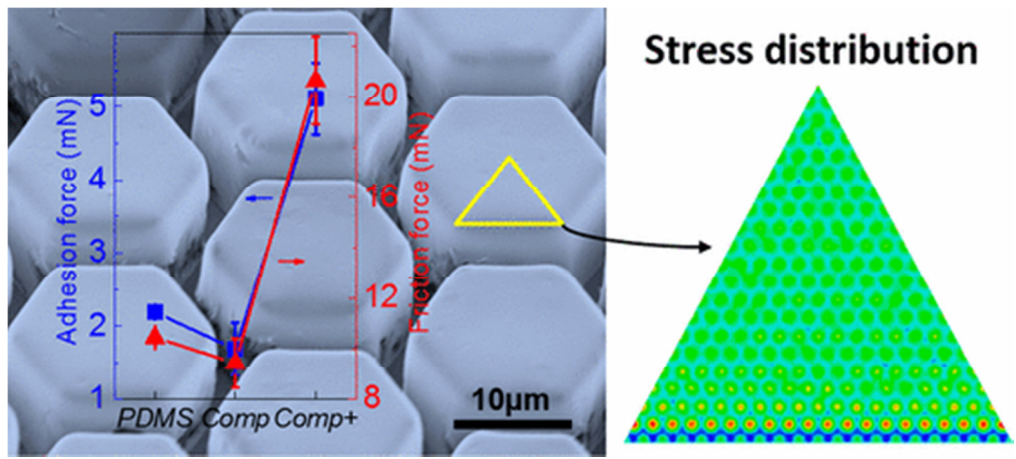

**Abstract.** Biological materials achieve directional reinforcement with oriented assemblies of anisotropic building blocks. One such example is the nanocomposite structure of keratinized epithelium on the toe pad of tree frogs, in which hexagonal arrays of (soft) epithelial cells are crossed by densely packed and oriented (hard) keratin nanofibrils. Here, a method is established to fabricate arrays of tree-frog-inspired composite micropatterns composed of polydimethylsiloxane (PDMS) micropillars embedded with polystyrene (PS) nanopillars. Adhesive and frictional studies of these synthetic materials reveal a benefit of the hierarchical and anisotropic design for both adhesion and friction, in particular, at high matrix–fiber interfacial strengths. The presence of PS nanopillars alters the stress distribution at the contact interface of micropillars and therefore enhances the adhesion and friction of the composite micropattern. The results suggest a design principle for bioinspired structural adhesives, especially for wet environments.

Biological materials often contain anisotropic building blocks assembled along preferred orientations to achieve directional reinforcement.(1) Musculoskeletal tissue, wood, and mollusk shells are relevant examples. The directional assembly of micro- or nanocomponents (*e.g.*, collagen fibers, cellulose fibers, and inorganic platelets) provides structural anisotropy and directional mechanical properties in bulk natural materials. Directionality is also a relevant property in the design of natural surfaces, like keratinized or cornified epithelium. A relevant example is the composite structure of the adhesive pads in the digits of tree and rock frogs.(2-8) On the toe pads of tree frogs, hexagonal arrays of (soft) epithelial cells separated by narrow channels are crossed by densely packed (hard) keratin nanofibrils (Figure 1).(8) These keratin fibers are oriented at an angle between 60 and 90° relative to the toe pad surface.(6, 7) Previous reports with living animals(4-10) and artificial models(11-13) have highlighted the important role of the surface micropattern to achieve friction enhancement on humid and flooded surfaces. However, the benefit of the embedded directional nanofibers in the microcomposite structure remains unclear.

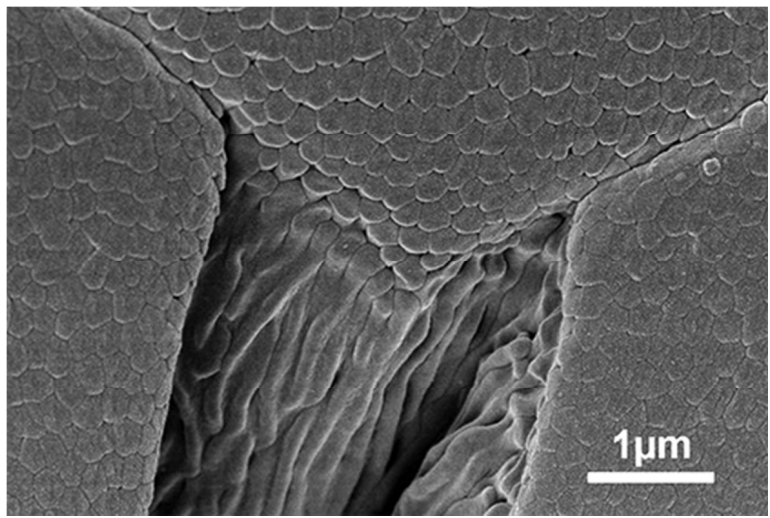

**Figure 1.** Scanning electron microscopy image of the edge of an epithelial cell on the toe pad of rock frog *Sphaerodactylus parvus*, showing a dense array of nanopillars covering the pad surface. Reproduced with permission from ref 8. Copyright 2015 Royal Society Publishing.

Bioinspired fibrillar adhesives for attachment in dry environments (gecko-like) have been studied over the past decade using artificial micro- and nanostructured models.(14-22) Arrays of closely packed nanofibrils are beneficial for conformability to real, rough surfaces, making the surface apparently "soft" even if it is composed of hard fibers that are resistant to mechanical damage and wear during locomotion.(18, 19) When a large amount of liquid is present at the contact interface, fibrillar surfaces are no longer effective for achieving high adhesion and friction, unless the liquid is drained from the interface by application of shear forces forming dry contacts.(8, 11-13) This working mechanism appears to be exploited by tree frogs for strong attachment during climbing on wet and flooded surfaces, but with a surface design different than that of geckos. Tree frog toe pads comprise composite microstructures combining a softer matrix (epithelial cells) and embedded hard keratin fibers.(6, 7) In general, a material with low elastic modulus is beneficial for strong adhesion but not necessarily for strong friction because of easy wear and damage.(23-29) However, both strong adhesion and friction are important for the jumping-based locomotion of tree frogs.(30)

The design of composite microstructures on the toe pads may contribute to the unique abilities of tree frogs. Fabrication of composites with controlled micro/nanostructures is challenging.

Surface patterns with aligned magnetic nanoparticles have been demonstrated.(31, 32) Hybrid patterns with silicon fibers embedded in a hydrogel matrix(33, 34) and vertically aligned carbon nanotubes embedded in polymeric materials(35-37) have also been realized. However, a composite micropillar array composed of perpendicularly oriented nanofibrils embedded in soft elastomeric matrix, mimicking the surface design of tree frogs, has not been realized. This is due, in part, to the difficulty of obtaining such composite micropillar arrays. Here, we report a fabrication method to obtain composite surface micropatterns with a soft elastomeric matrix and perpendicularly oriented polymeric nanopillars with tunable interfacial interactions, mimicking the toe pad structure of tree frogs. The composite surface structures are fabricated in a distinct pattern to realize enhanced adhesion and friction properties in a single system. Our patterns mimic the geometrical pattern and mechanical properties of keratinized epithelium of the tree frog's adhesive toe pads, but the reported method is generic and flexible and can be extended to other surface designs and material combinations.

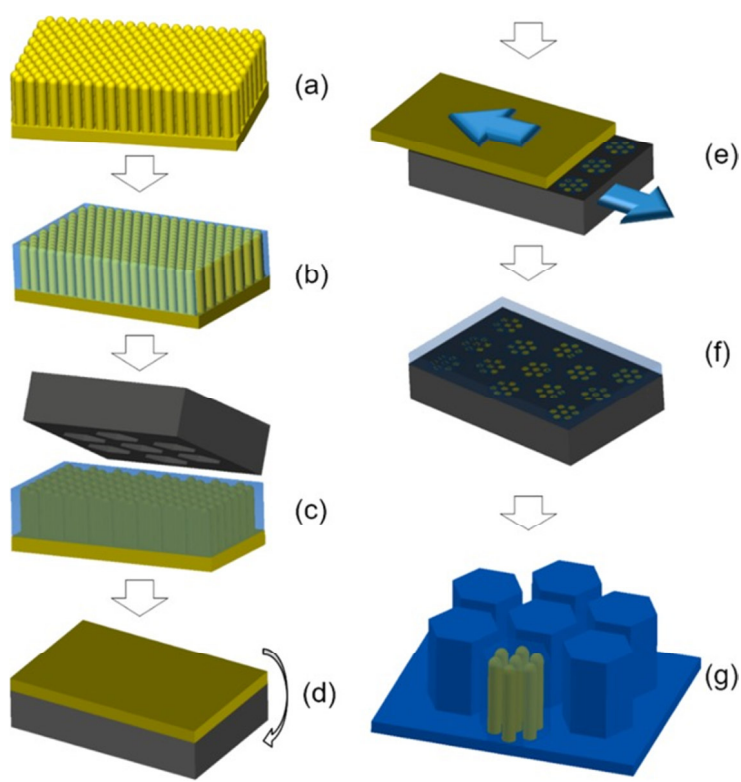

**Figure 2.** Schematic showing the fabrication steps of the composite micropillar patterns. (a) PS nanopillar array (yellow) that was replicated from an AAO template; (b) after optional PS surface modification with vinyl groups, casting of PDMS precursor (light blue) fills the nanopillar array and forms a thin film on top; (c) embossing with Ni mold insert (dark gray) on PS nanopillar array with precursor film at high pressure; (d) flipping over and cooling down in liquid $N_2$; (e) shearing and breaking of nanopillars from PS substrate in liquid $N_2$; (f) casting of thick PDMS precursor on the Ni mold insert to act as backing layer; (g) curing of PDMS backing layer and demolding resulted in composite pillar arrays (blue). One pillar shows the PS nanopillars inside.

## Results and Discussion

The composite micropatterns consist of arrays of soft polydimethylsiloxane (PDMS) micropillars (Young's modulus, $E \sim 2$ MPa) with rigid polystyrene (PS) nanopillars ($E \sim 3$ GPa) embedded in the micropillars and oriented perpendicular to the surface. This hierarchical composite micropattern was obtained via a multistep process (Figure 2, experimental part). First, hexagonal arrays of cylindrical PS nanopillars with a period of 500 nm, rod diameters of 330 nm, and a rod height of 10 μm (Figure 3a) were obtained by replicating from anodic aluminum oxide (AAO) membranes.(19) The following two treatments of the PS nanopillar array are very important to achieve the designed tree-frog-inspired structure: (1) The surface of PS nanopillars was modified with vinyl groups in order to covalently link the nanopillars to the PDMS matrix. (2) The gaps within the PS nanopillar array were completely filled with PDMS precursor to prevent the possible fusing of PS nanopillars in the shearing step.

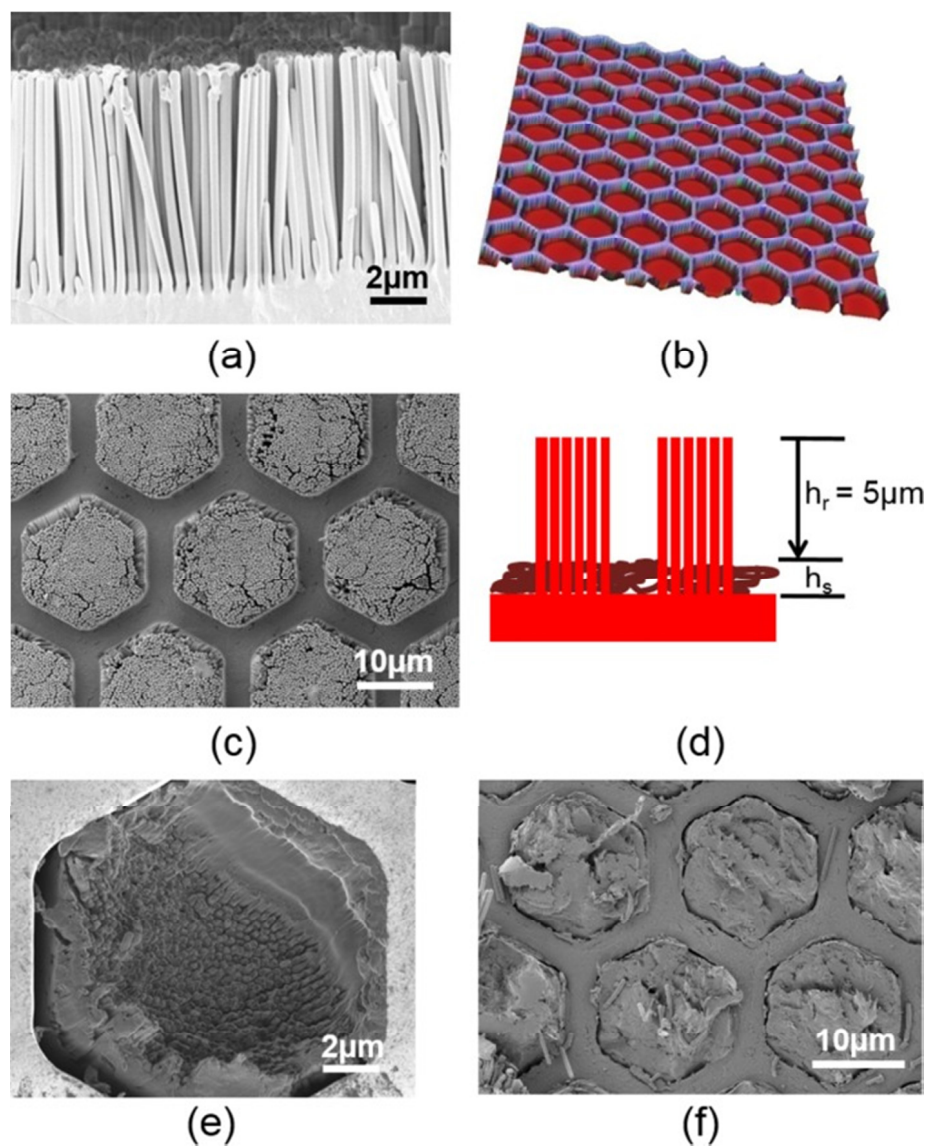

**Figure 3.** PS nanopillar arrays. (a) Scanning electron microscopy (SEM) image of the cross section of the PS nanopillar array. (b) Confocal microscopy image of the pattern on the nickel mold insert. (c) SEM image of patterned PS nanopillar array. (d) Schematic drawing shows that the remaining height ($h_r$) of PS nanopillars (red) is designed to be 5 μm. SEM image of (e) footless nanopillars retained in the nickel stamp after shearing and (f) corresponding supporting layer.

The PS nanopillar/PDMS precursor sample was molded under pressure using a micropatterned Ni mold insert (Figure 3b) with the negative copy of the hexagonal micropillar pattern. This molding process crushed the PS nanopillars selectively below the walls of the Ni shim but not within the holes. In this way, a hexagonal microchannel pattern (channels *ca.* 5 μm deep and 3 μm wide) was superimposed onto the PS nanopillar/PDMS liquid array (Figures 3c and S1). The dimensional size of the PS nanopillar array and the pressure applied on Ni shim were chosen such that the height of the remaining pillar ($h_r$) is 5 μm (Figure 3d).

The bottoms of the channel between the pillars has a smooth surface, indicating the pressure was large enough to crush the nanopillars below the walls of the Ni shim; on the other hand, the standing configuration of remaining PS nanopillars suggests that the pressure was not too large to destroy the nanopillars within the holes.

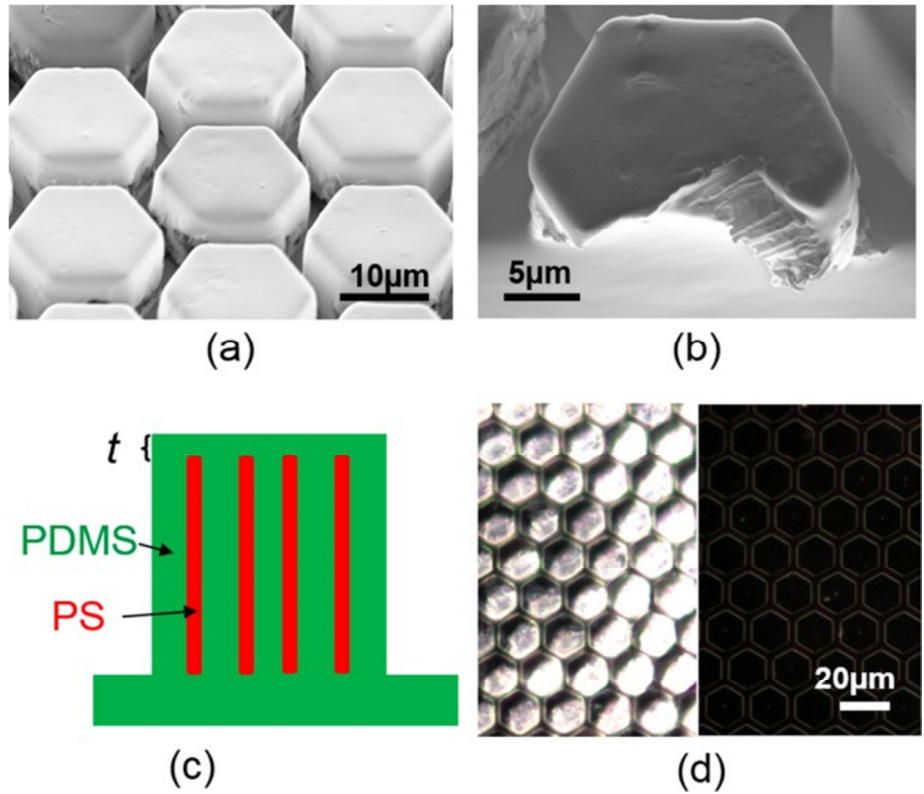

**Figure 4.** Resulting composite pillars with PS nanopillars embedded in the PDMS matrix. SEM image of (a) composite pillar array *Comp+* and (b) cross section of a *Comp+* pillar. (c) Schematic drawing shows that the thickness of PDMS layer ($t$) on top of PS nanopillars in the *Comp+* pillars. (d) Comparison of the dark-field microscopy image of *Comp+* (left) and PDMS (right) pillar arrays.

After shearing off the supporting layer of the PS nanopillar array, PS nanopillars remained embedded in the PDMS precursor, did not collapse, and retained a length of ~5 μm. Only a few nanopillars at the edges of the microstructures, close to the walls of the Ni template, deformed during shearing and collapsed with their neighbors (Figure 3e). The remaining supporting layer of the nanopillar array also showed a clear micropattern (Figure 3f), confirming an effective, clean fracture of the PS nanopillars and the robustness of the fabrication method. The sample was brought back to room temperature in a vacuum oven in order to avoid condensation of water on the surface. PDMS precursor was then added on top of the Ni mold insert and cured to form a backing layer supporting the composite microstructures. After curing of the PDMS, the array of composite micropillars on a PDMS layer was peeled off from the Ni mold. The resulting composite micropillar array is referred to as *Comp+* throughout this article. For comparison, PS nanopillars without the treatment of vinyl groups were also embedded into the PDMS matrix by using the same fabrication process. The composite pillars without the vinyl group are referred to as *Comp* in the following text.

Figure 4a,b shows the scanning electron microscope (SEM) image of the composite micropillar array. The pillars have a smooth PDMS top surface. A cross section of the pillars shows the embedded, standing subsurface nanopillars in a slightly tilted orientation as a consequence of the shearing process (Figures 3e and 4b). Considering the tilted configuration

of the nanopillars, the thickness of the top layer of PDMS ($t$) on *Comp+* was estimated to be ~100 nm (Figure 4c). Figure 4d shows the dark-field microscopy image of *Comp+* as well as the PDMS micropillar without embedded PS nanopillars. In the dark-field microscope, the existence of PS/PDMS interfaces caused the composite pillars *Comp+* (left side of Figure 4d) to be brighter than the PDMS pillars without embedded nanopillars (right side of Figure 4d). These results show that the proposed processing method allows fabrication of anisotropic, multicomponent micropatterns with embedded and aligned nanofibers. The method was optimized here for PDMS/PS material combination and a particular geometry in order to mimic the properties of tree frog attachment pads. However, the techniques are not PS- or PDMS-specific and could be extended to other material types, provided that one material can flow into nanopores (for the nanofibers) and the other can be cured after the molding step.

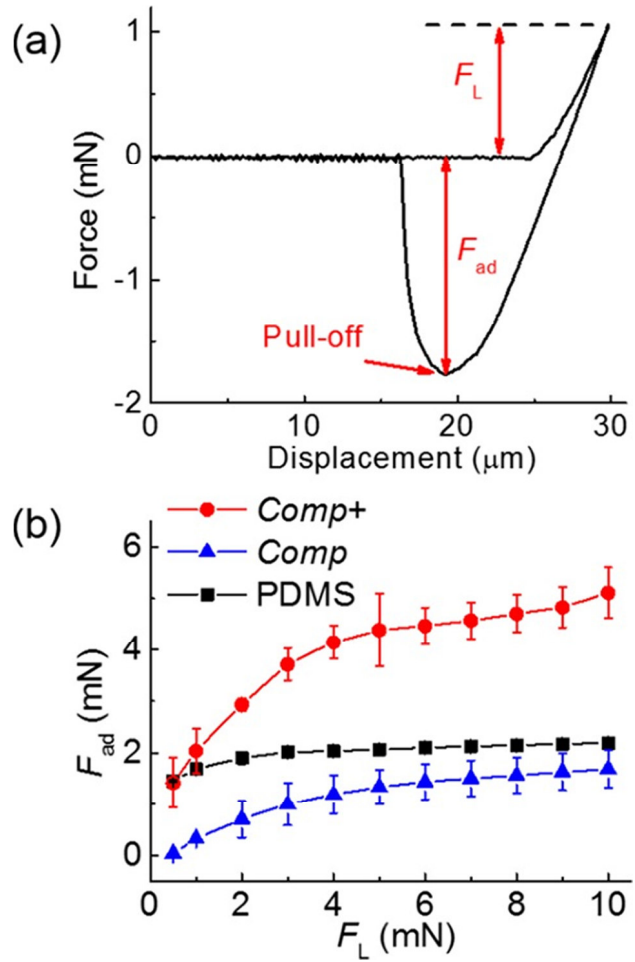

**Figure 5.** Evaluation of adhesion performance. (a) Representative force–displacement curve measured on PDMS pillar arrays. Loading force ($F_L$), adhesion force ($F_{ad}$), and the pull-off point are indicated. (b) Dependence of $F_{ad}$ on $F_L$ measured on PDMS, *Comp*, and *Comp+* micropillar arrays. Each data point in (b) represents the mean value of three measurements. Standard deviations are indicated by error bars.

The adhesion and friction behavior of the tree-frog-inspired composite micropillar arrays were characterized using a spherical ruby probe with a diameter of 5 mm.(11, 12) A typical force–displacement curve of the adhesion test is shown in Figure 5a, highlighting the value of the pull-off force ($F_{ad}$) as a measure of the adhesion performance. The adhesion forces at different loading forces (1 to 10 mN) were evaluated for the *Comp* and *Comp+* arrays and compared to the arrays of PDMS without embedded nanopillars.(12, 38) It should be mentioned that there is no liquid added to the contact interface in all of the tests reported here. In previous work, it has been shown that the hexagonal design can drain liquid out from the contact interface and form solid–solid direct contact on a wet surface where van der Waals forces may contribute mainly to the adhesion forces.(12, 13, 39) As the current hexagonal pattern design is identical to the previous one,(12) we assume the structure proposed here will have the same draining effect on a wet surface. Therefore, the current work focuses on the evaluation of adhesion and friction on dry surfaces. The experimental results show that the $F_{ad}$ of the *Comp+* arrays was significantly higher than $F_{ad}$ of a PDMS pattern without the PS reinforcement, whereas $F_{ad}$ on *Comp* arrays was lower than that of the PDMS pattern (Figure 5b). Both the presence of the hard subsurface nanopillars and the strength of the interface between the rods and the PDMS matrix influence adhesion performance.

Both the *Comp+* and *Comp* structures can be considered composite materials. Assuming perfect alignment of the PS nanopillars in PDMS, the effective elastic modulus of the composite pillar $E_{comp}$ can be estimated as $E_{comp} = V_{PS}E_{PS} + V_{PDMS}E_{PDMS}$, where $V$ is the volume fraction for each material and $E$ is the Young's modulus. Taking $E_{PS}$, $E_{PDMS}$, and $V_{PS}$ to be 3 GPa, 2 MPa, and 0.39, respectively, the estimated $E_{comp}$ is 1.2 GPa, which is 500 times higher than $E_{PDMS}$. In general, a material with smaller $E$ allows greater conformation to the contact surface and therefore can result in stronger adhesion.(40) Biological systems bearing adhesive toe pads, like geckos and tree frogs, employ hierarchical structures to reduce apparent $E$ and to enhance adhesion. For example, the gecko setae is composed of a hard material (β-keratin) with an $E$ of 2–4 GPa, whereas the apparent $E$ of the seta array is only around 100 kPa.(41) However, both enhanced and slightly reduced adhesions were found in the composite pillars (*Comp+* and *Comp*) with larger apparent $E$ as compared to the PDMS micropillars. Obviously, the simplistic argument in terms of the effective stiffness of the composite materials does not apply to our case. On the other hand, it is also quite clear that the rigid PS nanopillars in the PDMS matrix play a critical role in determining the adhesion performance of the composite pillars.

The stress distribution at the contact interface, which is influenced by both the tip geometry of pillars and the elastic heterogeneity of the bodies in contact, is critical to understand the adhesion enhancement observed in the *Comp+* samples. From studies related to gecko adhesion, it is known that mushroom-shaped tips on micro- and nanopillars alter the stress distribution and result in a maximum stress at the contact center, and that the stress decreases smoothly out to the contact perimeter.(42, 43) In general, it is more difficult to initiate a crack at the center than at the edge of a contact. Furthermore, the crack initiation on the pillar will start from the contact center and propagate toward the edge of the mushroom tip, which could introduce a vacuum pressure contribution to the pull-off force.(44) Therefore, these two effects (*i.e.*, the crack initiation from a stress center and the vacuum pressure) result in a large pull-off force (adhesion force). In contrast, the crack initiation on a simple cylindrical micropillar without an overhang structure happens at the pillar edge, where the maximum stress is located. A stress distribution with a high stress at the center relative to the edge has also been realized in the millimeter and sub-millimeter composite pillars that are composed of a stiff core ($E_{core}$ > 3 GPa) and a thin shell of PDMS, without the overhang structure.(45, 46) As the thickness of the PDMS layer on top of the stiff core is decreased, the maximum stress shifts to the center and the detachment force increases.(45)

Considering the strong interfacial bonding between the PS nanopillars and PDMS matrix (*Comp+*) in the structures investigated here, the PS/PDMS composite effectively acts as a stiff core, similar to the previous work discussed above. However, the discrete nature of the PS nanopillars may affect the stress distribution, thus we investigated our structure using finite element analysis (Figure 6). Due to the symmetric nature of the micropillars here, 1/6 of the micropillar (in the shape of equilateral triangle) was simulated (Figure S2a) in the primary simulation. In a separate simulation, a representative cell containing two quarters of PS nanopillars and the matrix (Figure S2b) was also simulated. The models were 3D and included the thickness of the PDMS layer ($t$) on top of the PS nanopillar, which was estimated to be around $t = 100$ nm (Figure 4c). All of the simulation results here are presented in terms of the local normal stress at the interface divided by the average normal stress at the interface. From the representative cell simulations (the black line represents the perimeter of the PS nanopillars), a local stress maximum is found to be located at the center of nanopillars (Figure 6a). The maximum normalized stress is 1.39, and the stress decreases smoothly toward the nanopillar perimeter. It should be noted that the normalized stress at the nanopillar perimeter is between 1.10 and 1.16 and decays further beyond the area of the nanopillar. It suggests that the strong interfacial bonding effectively transmits stress between the PS nanopillar and the PDMS matrix.

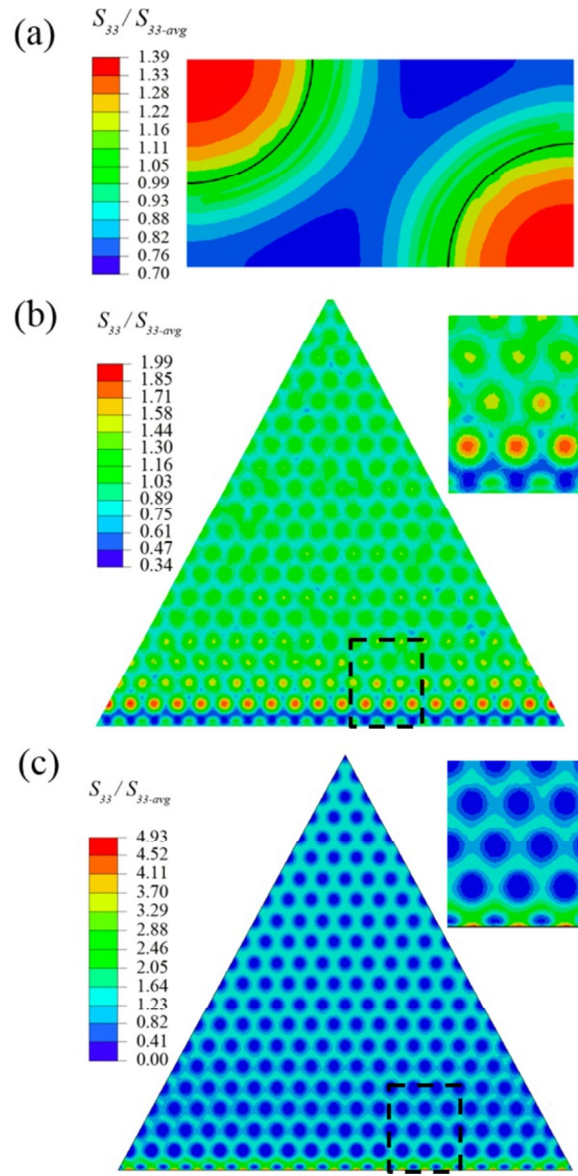

**Figure 6.** Simulated stress distribution on *Comp+* micropillars during detachment. (a) Representative cell of the composite pillar containing two quarters PS nanopillars. The black line indicates the edge of PS nanopillar. Stress distribution on the (b) *Comp+* and (c) *Comp* micropillar. The insets in (b,c) are the zoomed-in views of the corresponding area in the dashed boxes in (b,c).

The stress distribution across the entire hexagonal composite pillar is quite complex (Figure 6b). Globally, the stress distribution appears quite homogeneous, especially in the central region. In the central region, the stress is locally high above the nanopillars and then decays in the area between the nanopillars, just as that in the representative cell. However, the global stress minimum (0.34) is found at the edge of the hexagonal composite pillar (at the bottom edge of the triangle in the 1/6 model, inset in Figure 6b). The maximum stress (1.99) is located on one to two rows of PS nanopillars and is just some distance away from the region where the stress minimum is located. It should be emphasized that the stress maximum is not at the edge of the micropillar. Furthermore, the stress maximum on the composite pillar is much smaller than the maximum normalized stress predicted along the edge of the pure hexagonal PDMS micropillar (2.58) (Figure S2c). This result clearly demonstrates that the presence of rigid nanopillars embedded in soft micropillar can both (1) reduce the value of stress maximum and (2) shift the stress maximum toward the central region (Figure 6b). It suggests that this kind of composite design in tree frogs may have a similar function to the overhang structures at the tip of seta in various animals bearing fibrillar adhesives.

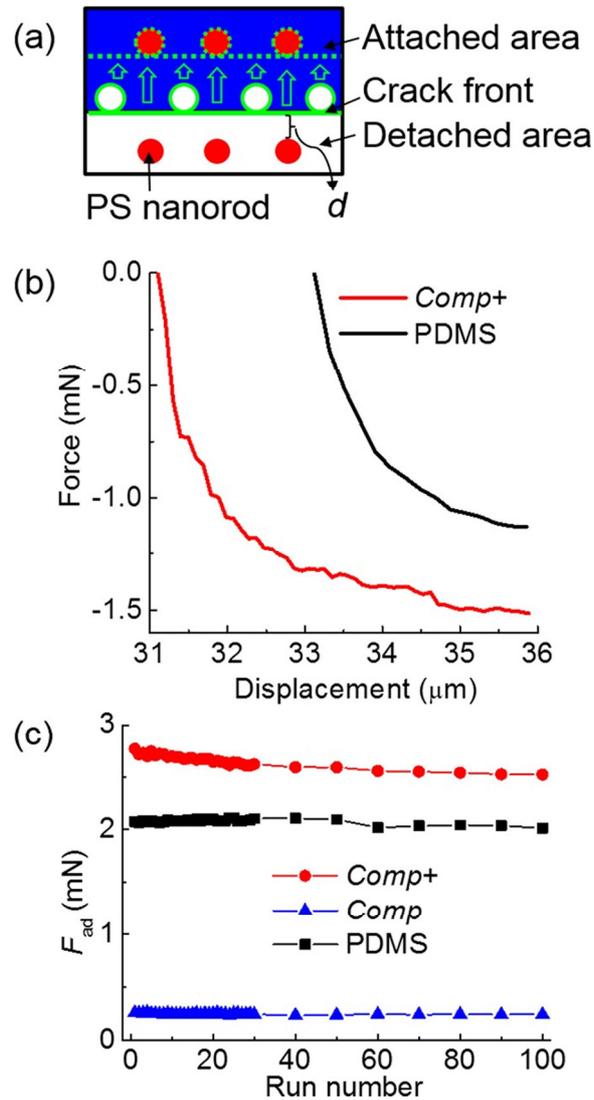

**Figure 7.** Detachment behavior and the structure stability. (a) Proposed movement of crack front during the detachment. The solid green line indicates the crack front; the dashed green line indicates the crack front at the next row of PS nanopillars; the green arrows indicate the moving direction of the crack front. (b) Fraction of the detachment curve after the pull-off point on *Comp+* and PDMS micropillar arrays. (c) Repeated adhesion tests on *Comp+*, *Comp*, and PDMS micropillar arrays.

When the interfacial bonding between the PS nanopillars and the PDMS matrix (*Comp*) is weak, the stress distribution is significantly different than that observed for *Comp+* (Figure 6c). As there is no strong chemical bonding between PS nanopillars and the PDMS matrix, the stress from the PDMS region, which directly contacts the contacting surface, could not effectively transmit to the embedded PS nanopillars. Therefore, the regions with lowest stress are located on top of the nanopillars. The stress within the PDMS matrix is rather homogeneous and just slightly higher than the stress minimum. The stress maximum is located at the micropillar edge within the PDMS area (inset in Figure 6c). Furthermore, the normalized stress reaches a maximum of 4.93, much higher than that of the pure PDMS pillar. This may explain why the *Comp* pillars have lower adhesion than the pure PDMS pillars (Figure 5b).

Once a crack initiates at the location of the maximum stress, the crack will propagate and be affected by the nonuniform stress distribution and elastic heterogeneity. The staggered arrangement of the PS nanopillars can further hinder the propagation of the crack front (green line in Figure 7a). The detaching part of the pull-off curve confirmed this discontinuity of crack propagation (Figure 7b). The detachment curve on *Comp+* showed a stepwise profile with periodic spacing between steps of 114.7 ± 3.8 nm, which corresponds to the shortest distance between rows of PS nanopillars of $d$ = 103 nm (Figure 7a). In contrast, the retract curve on the pure PDMS pillar has a smooth profile.

This phenomenon is similar to the detachment of an adhesive layer with subsurface microchannels filled with different liquids.(47, 48) The crack front does not propagate continuously at the interface. The crack is arrested close to the location with minimum modulus and only initiates again at larger peel-off force. The discontinuous crack propagation therefore results in a higher adhesion compared to a uniform material. In previous work,(47, 48) modulation of the stiffness of the PDMS was achieved by filling the buried microchannels with a fluid (air or liquid), which cannot very well sustain stresses. In our case, the second component (PS nanopillar) is a hard solid and links to the PDMS matrix *via* chemical bonding, which can transfer the stress as demonstrated by finite element analysis (Figure 6a). Furthermore, no obvious change in interfacial interaction and a high efficiency of deformation across the interface between two materials can be concluded from the 100 cycles of attachment/detachment at the same location because the adhesion performance kept constant (Figure 7c).

Figure 8a shows the friction curve measured on the composite structure of *Comp+*.(12, 28, 29) The friction forces of the PDMS pillars show a linear dependence on the normal loading force and a friction coefficient of ∼0.89. The friction performance on composite pillar *Comp* was identical to that on the PDMS micropillar surface (Figure 8b). However, *Comp+* micropatterns showed significantly higher friction. Under a normal load of 1 mN, *Comp+* showed a 88% friction enhancement. During friction, the micropillars tilt and elongate along the shear direction. The deformation of the PDMS pillar and the adhesion force between the pillar and the probe contribute to the friction force. The shearing force of one pillar, $F_{shear}$, can be estimated by $F_{shear} = GA\Delta x/h$, where $G$, $A$, $h$, and $\Delta x$ are the shear modulus, the area of the pillar top, the height of the pillar, and the transverse displacement of the pillar along the shear direction, respectively.(49) The sum of $\Delta x$ of the pillars in contact is the displacement at which static friction changes to kinetic friction in the friction curves, $D_s$ (Figure 8a). Therefore, the apparent $F_{shear}$ can be estimated from $F_{shear} \sim nGAD_s/h$, where $n$ is the number of pillars in contact. The stronger adhesion on *Comp+* micropatterns leads to larger lateral displacement needed to initiate the detachment of the pillar edge at the shearing front. In fact, $D_s$ on *Comp+* doubles that on the pure PDMS pillar (Figure 8c). *Comp+* micropatterns also show higher shear stiffness, which may facilitate transfer of shear stress to the PDMS backing layer, which can ultimately result in a larger deformed volume and increased dissipation. Both factors together resulted in a larger friction force for our composite patterns.

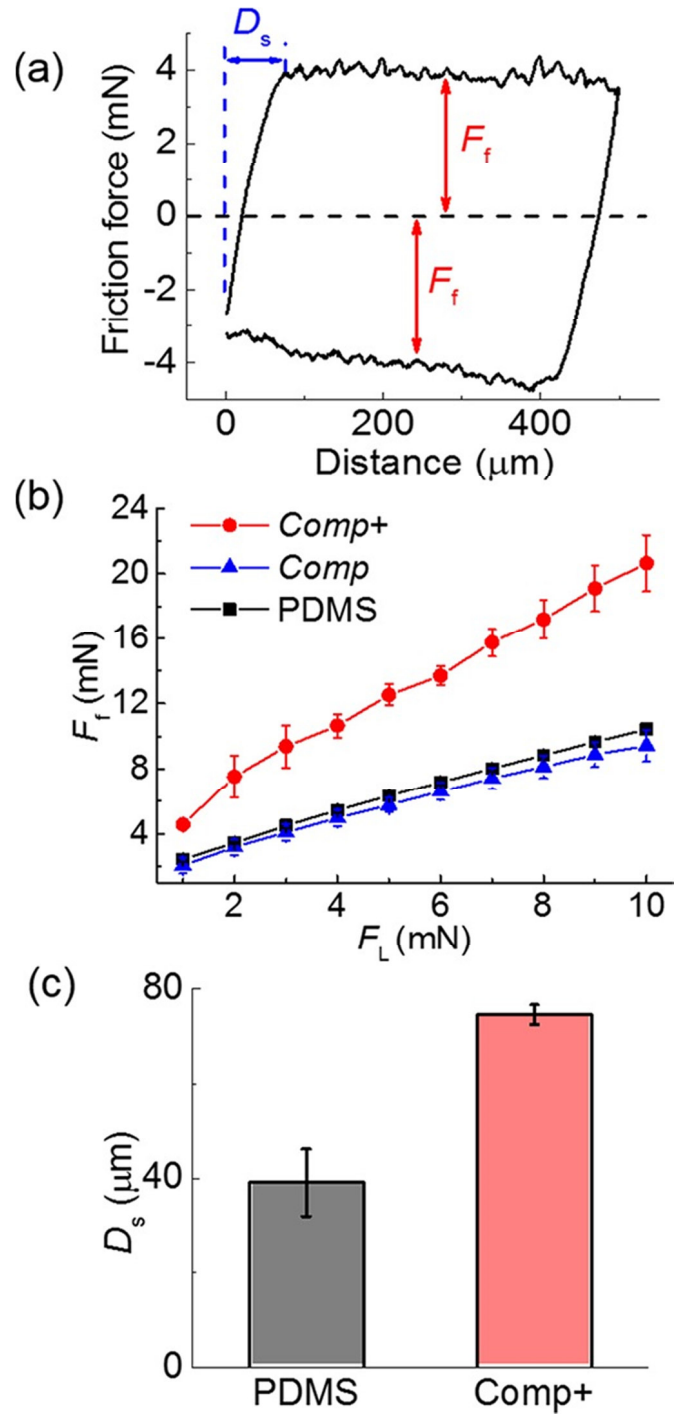

**Figure 8.** Evaluation of friction performance. (a) Representative friction curve measured on PDMS arrays. The friction force ($F_f$) in trace and retrace directions and the transition distance from static to dynamic friction, $D_s$, are indicated. (b) Dependence of $F_f$ on $F_L$ on *Comp+*, *Comp*, and PDMS micropillar arrays. (c) $D_s$ measured on *Comp+* and PDMS micropillar arrays. Each data point in (b,c) represents the mean value of three measurements. Standard deviations are indicated.

In order to verify the importance of the freestanding, embedded nanopillars for the enhanced adhesion and friction performance on *Comp+* patterns, two control experiments were carried out: one on a flat composite structure consisting of a PDMS film with an embedded PS sheet, and one on a microstructured PDMS with embedded nanofibers linked to their stiff PS backing layer (Figure S3). The friction experiment on the flat PDMS film with the embedded PS sheet caused significant damage of the PDMS surface (Figure S3a). The friction curve showed a large static friction force, followed by a drop in friction after surface damage (Figure S3b), and a weaker dynamic friction associated with the pushing of the PDMS layer along the rigid PS layer (Figure S3a). The friction damage was partially reduced when the PS film was covalently bound to the PDMS (Figure S3c). When embedded nanopillars were connected to the PS backing layer, damage of the top PDMS layer was also observed, in contrast to the stability of *Comp+* micropatterns, where the PS nanopillars were not connected to a stiff backing layer (Figure S3d). The presence of a rigid, continuous layer underneath the soft PDMS top layer will lead to highly localized stresses that result in damage of the top layer. Thus, our *Comp+* structures offer a combination of several mechanical properties: (i) modulation of the local effective Young's modulus due to the stiffness contrast between PDMS and PS leading to improved adhesion by crack arrest; (ii) homogeneous distribution of the stress on the pillar top and efficient transfer of stress from the pillar top to the backing PDMS layers; and (iii) the delicate balance between lateral bending stiffness of the pillars and flexibility of the top PDMS layer to increase compliance to a counterpart surface.

# Conclusions

We developed a method to fabricate composite micropillar patterns reinforced with hard, rootless nanopillars. We applied our approach to the PDMS/PS system in order to mimic the design of tree frog's adhesive toe pads. However, the method could be extended to other polymers or material combinations. Enhanced adhesion and friction were found on composite pillar arrays *Comp+*, where the PS nanopillars and the PDMS were covalently linked, allowing transmission of mechanical stress and deformation. These results suggest that the hierarchical structure found in the surface of tree frog attachment pads is beneficial for both adhesion and friction and possibly required for tree frog's survival. Combining the composite design of the microstructure and the presence of liquid at the contact interface may deepen our understanding of the adhesion abilities of tree frogs, which will come in a subsequent paper. The results here also provide insight for the design of bioinspired materials with both strong adhesion and friction based on composite structures without the complicated fiber geometries typically used in gecko-inspired dry adhesives.

# Experimental Section

**Materials.** Polydimethylsiloxane elastomer kits (Sylgard 184) were purchased from Dow Corning (MI, USA). Polystyrene ($M_w$ = 35 kg mol$^{-1}$; PDI = 1.04) and vinyltriethoxylsilane (analytical grade) purchased from Sigma-Aldrich were used as received. Self-ordered AAO templates were fabricated by two-step mild anodization with phosphoric acid as electrolyte.(50) The as-prepared AAO templates had a pore diameter of ∼180 nm, a lattice period of ∼500 nm, a pore depth of 10 μm, and a round area with a diameter of 15 mm. The pores were widened in 10% phosphoric acid at 30 °C for 65 min. The widened AAO templates were washed and dried in vacuum.

**Equipment.** Surface microstructures were characterized by white light confocal microscopy (μsurf, Nanofocus AG, Oberhausen, Germany) and scanning electron microscopy (LEO 1530VP Gemini; Carl Zeiss Jena, Oberkochen, Germany). The surface modification was carried out on a Plasma Activate Statuo 10 USB (Plasma Technology GmbH, Rottenburg, Germany).

**Fabrication of the Microstructured Nickel Mold Inserts.** For the mold insert fabrication, a Cr/Au metallized 2 in. silicon wafer (which contains an etched pattern field of 10 × 10 mm$^2$) was used as master. The Si master was fixed on a poly(methyl methacrylate) sheet by adhesive tape and contacted by a copper wire. Nickel electroforming was carried out in a boric acid containing nickel sulfamate electrolyte (pH 3.4–3.6 at 52 °C) for approximately 48 h. To ensure a slow growth of the nickel layer at the beginning and to achieve a defect-free galvanic filling of the microstructures, the current density was adjusted to 0.1 A/dm$^2$ (corresponding to a growth speed of approximately 0.02 μm/min) at the start of the plating process. After every 30 min, the current density was increased from 0.1 up to 1.5 A/dm$^2$ (approximately 0.3 μm/min). If the desired metal thickness (500 μm) was reached, the silicon wafer with the thick nickel layer was dismounted and the silicon wafer was removed by wet-chemical dissolving using 30% KOH. After a plasma stripping and final cleaning procedure with isopropyl alcohol shaking for 10 min, the mold insert was usable for further SEM characterization (Figure S4). The dimensions of the hexagonal holes (and therefore the replicated micropillars) were 20 μm in diameter (*D*), 5 μm in height (*H*), and 3 μm in gap width (*W*) between the pillars following the design of a frog's adhesive toe pad.(11, 12)

**Preparation of the PS Nanopillar Array.** The preparation of the PS nanopillar array followed the procedure previously reported elsewhere.(19) In a typical procedure, the PS film was placed on AAO (Figure S5) and heated to 200 °C for 3 h under vacuum while a pressure of about 160 bar was applied. Aluminum layer in AAO template was dissolved by immersion in a solution of 100 mL of 37% HCl and 3.4 g of $CuCl_2 \cdot 2H_2O$ in 100 mL of deionized water at 0 °C. After the removal of the aluminum layer, the alumina layer was etched away in 1 M aqueous NaOH solution at room temperature for 1 h, and the NaOH solution was replaced with a fresh one for another 1 h. The PS sample was then dried in a freeze-dryer to avoid collapse of the nanopillars by capillary forces during drying (Figure 2a). A typical PS nanopillar array has a dimension of 5 × 5 mm$^2$.

**Surface Modification of the Nanopillar Array.** The PS nanopillar film was adhered to a PS plate with superglue. PS nanopillars were treated with oxygen plasma with 100 W, 0.1 mbar for 30 s. The treated PS nanopillars were immersed in solution of vinyltriethoxylsilane dispersed in mixed $NH_3/H_2O$/ethanol for 30 min. The treated sample was freeze-dried.

**Removal of the Supporting Layer of the PS Nanopillar Array.** Approximately 10 μL of PDMS precursor was casted on the vinyl-modified PS nanopillar array (Figure 2b). The thickness of the PDMS layer was just enough to cover the PS nanopillars. The PS nanopillar arrays were brought into contact with the nickel mold insert (Figure 2c). The assembly of the PS nanopillar array and the nickel insert was mounted on the cantilever of a homemade shearing device with strong magnets, as shown in Figure S3. An extra pressure was added to the magnets for 1 min. The assembly of the PS nanopillar array/Ni shim was then immersed into liquid nitrogen for *ca.* 30 s for deep-freezing (Figure 2d). At this temperature, the PDMS precursor becomes glassy and the PS nanopillars are brittle. By rotating the screw on the lower manipulator (Figure S6a), the upper manipulator was moved rightward to apply a shear force on the PS substrate (Figures 2e and S6b). A moving distance of around 5 mm allows PS nanopillars (remaining in Ni mold) to be completely sheared off from the PS backing layer. The whole device was then heated up to room temperature in a vacuum oven. Afterward, the PS plate with the supporting layer was removed, and the PS nanopillars embedded in PDMS remained in the nickel mold insert.

**Preparation of Composite Pillars.** PDMS precursor was cast onto the nickel insert and left at room temperature for 30 min to form a backing layer (Figure 2f). This period had two functions: (1) allowing the PDMS precursor to fill any remaining voids inside the nanopillar array in the nickel insert; (2) leveling off the free surface of the PDMS backing layer. The sample was then heated to 60 °C for 4 h. The composite pillar was then ready to peel off from the nickel insert (Figure 2g) and is denoted as *Comp+* throughout the paper. For the comparison, composite pillars composed of PDMS matrix and embedded PS nanopillars without the treatment of vinyl groups were also fabricated with the same dimensions and curing history, which are referred to as *Comp*.

**Preparation of Pillar Arrays with PDMS.** The thermal treatment for curing PDMS was the same for both *Comp+* and *Comp*.

**Adhesion and Friction Tests.** Adhesion and friction tests were all carried out on a homemade device (PIA) as shown in our previous work.(11, 12, 28) In brief, a spherical ruby probe of 5 mm in diameter is connected to the upper force sensor, which controls the loading forces in adhesion and friction tests. The sample was mounted on the lower sensor, which records the lateral friction force. All the tests were performed in ambient conditions without any liquid at the contacting interface. In adhesion tests, the sample surface approached the probe at a speed of 20 μm/s until a predefined loading force was reached. The sample was then retracted at the same speed. The adhesion force corresponds to the value of the force at the pull-off event.

In friction tests, the probe was brought into contact with the sample surface and a normal force was applied and kept constant during lateral shearing. The sample was moved at a velocity of 100 μm/s over a distance of 500 μm, forward and backward, while the forces were simultaneously recorded.

# Supporting Information

The Supporting Information is available free of charge on the ACS Publications website at DOI: 10.1021/acsnano.7b04994.

# Acknowledgment

We thank Prof. Jon Barnes (University Glasgow) for fruitful discussions, Christian Benkel at KIT for the preparation of the nickel shim. L.X. thanks the National Natural Science Foundation of China (51503156, 51611530546) and "Young 1000 Talents" for support. A.d.C. thanks the Deutsche Forschungsgemeinschaft within the program SPP1420 "Biomimetic Materials Research: Functionality by Hierarchical Structuring of Materials" (Projects CA880/1, BU 1556/26), ITN BioSmarTrainee (Project No. 642861), and the Marie Sklodowska-Curie Innovative Training School, No. 642861, for financial support. M.S. thanks the European Research Council (ERC-CoG-2014; project 646742 INCANA) for funding. The fabrication of the Ni mold insert was kindly supported from the Karlsruhe Nano and Micro Facility (www.kit.edu/knmf), a Helmholtz Research Infrastructure at Karlsruhe Institute of Technology. K.T.T. and A.L. acknowledge the National Science Foundation (CMMI-1435745 and CMMI-1663037) for financial support for the finite element simulations.